\documentclass{elsart}
\usepackage{amssymb}

\usepackage{graphicx}
\usepackage{amsmath}

\begin{document}

\begin{frontmatter}
\title{Lorentz Invariance Violation and the QED formation length}
\author{H. Vankov}  and
\author{T. Stanev}
\address{
Institute for Nuclear Research and Nuclear Energy, Sofia, Bulgaria}
\address{
Bartol Research Institute, University of Delaware, Newark DE 19716, USA}

\begin{abstract}
It was recently suggested that possible small violations of Lorentz
invariance could explain the existence of UHECR beyond the GZK cutoff and
the observations of multi-TeV gamma-rays from Mkn 501. Our analysis of
Lorentz-violating kinematics shows that in addition to the modified
threshold conditions solving cosmic ray puzzles we should expect a strong
suppression of electromagnetic processes like bremsstrahlung and pair
creation. This leads to drastic effects in electron-photon cascade
development in the atmosphere and in detectors.

\end{abstract}
\end{frontmatter}

\section{Introduction}

 A tiny Lorentz invariance violation (LIV) at high energies was suggested 
 \cite{Coleman} as an explanation of two
 experimental astrophysical paradoxes - the observations of ultra high energy
 cosmic rays (UHECR) well beyond the theoretically expected
 Greisen-Zatzepin-Kuzmin (GZK) cutoff around $6\times 10^{19}$ eV \cite{GZK}
 and the observations of 20 TeV gamma-rays from Mkn 501. UHECR should have
 been absorbed in photoproduction collisions with the microwave background
 and the 20 TeV $\gamma $--rays -- on the extragalactic infrared/optical
 background~ \cite{Protheroe}. While some authors \cite{Aharonian,Berezinsky}
 consider the second puzzle not so dramatic and LIV hypothesis too premature,
 the number of showers above $10^{20}$ eV is already big enough to suggest
 the existence of a problem. As a solution of this problem LIV was first
 suggested about 30 years ago \cite{Kir,Sato} and later in \cite{Mestres}.
 In \cite{Coleman}, \cite{Amelino}, \cite{Kifune}, \cite{Aloisio} the LIV
 hypothesis was suggested as a solution for both paradoxes.

 In Ref.~\cite{Amelino} the Lorentz invariance violation was 
 formulated to correspond to 
 an energy dependent photon group velocity 
\begin{equation*}
\frac{\partial E}{\partial k}=c[1-\xi _{\gamma }\frac{E}{E_{0}}]\;.
\end{equation*}
Here $c$ is the speed of light. This corresponds to a dispersion relation 
\begin{equation}
c^{2}k^{2}\simeq E^{2}+\xi _{\gamma }\frac{E^{3}}{E_{0}}
\end{equation}
with $\xi _{\gamma }=\pm 1$ and $E_{0}\sim 10^{19}$ GeV.

 Kifune \cite{Kifune} has used this formulation to investigate the
 consequences of \ LIV on collisions of high energy radiation with soft
 photons. The detection of TeV photons from point sources and protons above
 the GZK cutoff sets some constraints on
 $\xi _{\gamma },\xi _{e}$ and $\xi_{p}$,
 when photons, electrons and protons are allowed to have different
 degrees of the LIV. He also noted that the modified relation of energy
 and momentum can affect mildly the detection of high energy radiation.
 Observations of $\gamma $-rays are based on pair creation in the detector
 material (satellite experiments) or on the detection of Cherenkov
 light from air
 showers initiated by TeV photons in ground-based observations.
 There are also numerous experiments using electromagnetic cascading
 in emulsion chambers to detect high energy electrons and $\gamma$-rays. 

 The common wisdom is~\cite{Amelino} that 
 although LIV affects significantly the GZK and TeV-$\gamma $
 thresholds the effect of the modified dispersion
 relation on other interactions of the relevant high-energy
 particles is negligible.

 The aim of the present paper is to discuss the possibility that
 LIV deformed dispersion relations could change not only the
 thresholds of some reactions at extremely high energies, but will
 also strongly affect the electromagnetic cascade development in
 the atmosphere and detectors. The reason for this is that deformed
 dispersion relations affect the \emph{\ formation length }of
 bremsstrahlung and pair production and, hence,
 their cross sections.

\section{Formation length and LI violation}

The longitudinal momentum transfer between highly relativistic interacting
particles (photons or electrons) and the target nuclei is small.
Ter-Mikaelian~\cite{Mikaelian} first realized that according to the
uncertainty principle the interactions take place not at a single point but
over a long distance (\emph{formation length, coherence length}). The
interactions cannot be localized within this length. In the classical
electrodynamics \emph{coherent length} is the distance over which
constructive interference between radiated waves takes place. During the
time the electron travels the formation length it and the radiated
electromagnetic wave separate enough~\cite{Shulga} to be considered
independent particles. In bremsstrahlung this separation is at least a
distance of the order of the emitted photon wavelength $\lambda $. In the
case of pair production the formation length is the length over which the
electron and the positron separate by a distance of about two electron
Compton wavelengths $2/m$.


While the transverse momentum exchanged with the nucleus, $q_{\perp },$ is
of order $mc$, the longitudinal momentum transfer $q_{\shortparallel }$ is
small. In bremsstrahlung 
\begin{equation}
\ q_{\shortparallel }=p_{e}-p_{e}^{\shortmid }-k/c \; ,
\end{equation}
where $p_{e}$ and $p_{e}^{\prime }$ are the electron momenta before and
after the interaction and $k$ is the photon energy. At high energy $E\gg m_e
c^{2}$ we can neglect emission angles and simplify to $q_{\shortparallel
}\sim \frac{m^{2}c^{3}k}{2E(E-k)}$. In pair production $q_{\shortparallel
}\sim \frac{m^{2}c^{3}k}{2E(k-E)}$. For bremsstrahlung the formation length
thus is: 
\begin{equation}
l_{0}\sim \frac{\hbar }{q_{\shortparallel }}= \frac{2\hslash E(E-k)}{%
m^{2}c^{3}k} \; .
\end{equation}
The formation length increases rapidly with the energy of the primary
particle and with its ratio to the energy of the emitted photon. For pair
production $E - k$ in the numerator is replaced by $k-E$.

The formation length is closely related to the interaction cross section.
The amplitude of the radiation is proportional to $l_{0}$, and its intensity
is proportional to $\sim l_{0}^{2}$. If there is emission from an electron
traversing a distance $D$, this is equivalent to $\frac{D}{l_{0}}$
independent emitters, giving a total radiation intensity proportional to $%
\mid l_{0}\mid ^{2}D/l_{0}\sim l_{0.}$~\cite{Klein}.

Because of the small value of $q_\shortparallel$ the formation length $l_0$
can have macroscopic dimensions even at moderate energies. For example, for
a 25 GeV electron, emitting a 100 MeV photon, $q_{\shortparallel }\sim 0.03$
eV/c and $l_{0}\simeq 10$ $\mu m.$ For 10$^{9}$ GeV electron and 10$^{5}$
GeV bremsstrahlung photon $q_{\shortparallel }\sim 1.45\times 10^{-8}$ eV/c
and $l_{0}\sim 14$ m.

The coherence over this very long length can be disrupted by other
interactions reducing the effective formation length and hence the
probability for radiation. For instance, because at high energies $l_{0}$
becomes much longer than the average distance between the atoms of the
medium, the multiple scattering of the electron on the atoms within the
formation length will change the electron path by reducing the electron
longitudinal velocity and the emission will be suppressed. This is the
physical mechanism of the Landau-Pomeranchuk-Migdal (LPM) effect (see \cite
{Klein} and references therein). Other factors that could reduce $l_{0}$ ,
and hence the probability for radiation or pair production are \cite{Klein}
photon interaction with the medium (dielectric suppression), external
magnetic field (magnetic suppression), suppression of bremsstrahlung by pair
production, and vice-versa.

In the case of bremsstrahlung the multiple scattering on the formation
length (LPM effect) leads to an additional term in the expression for $%
q_{\shortparallel }$ which increases the longitudinal momentum transfer. In
the small angle approximation 
\begin{equation}
q_{\shortparallel }\simeq \frac{m^{2}c^{3}k}{2E(E-k)}+ \frac{k\theta
_{MS}^{2}}{2c}\;,
\end{equation}
where $\theta _{MS}$ is the electron multiple scattering angle in half the
formation length \cite{Klein}. The increase of $q_{\shortparallel }$ reduces
the formation length. The multiple scattering becomes significant when the
second term in (4) is larger than the first.

LIV modified dispersion relation acts like the suppression factors above.
Let us put, similarly to Eq.~1, the square of the modified momentum 
\begin{equation}
q^{2}=p^{2}+\xi \frac{E^{3}}{E_{0}c^{2}}
\end{equation}
Then Eq.~2 becomes ($q\simeq p+\xi \frac{E^{2}}{2E_{0}c}$) 
\begin{equation}
q_{\shortparallel }\simeq \frac{m^{2}c^{3}k}{2E(E-k)}+\xi _{e}\frac{E^{2}}{%
2E_{0}c}-\xi _{e}\frac{(E-k)^{2}}{2E_{0}c}-\xi _{\gamma }\frac{k^{2}}{2E_{0}c%
}\ 
\end{equation}
for bremsstrahlung, and 
\begin{equation}
\ q_{\shortparallel }\simeq \frac{m^{2}c^{3}k}{2E(k-E)}+\xi _{\gamma }\frac{%
k^{2}}{2E_{0}c}
\end{equation}
for pair production. In Eq.~7 LIV dispersion relation is used only for the
particle with the highest energy - the photon.

The effect depends on the signs of the parameters $\xi $. If we take the
electron and positron energies in Eq.~7 equal to $\frac{k}{2}$ ($k$ is the
photon energy), then 
\begin{equation*}
q_{\shortparallel }^{\min }\simeq \frac{2m^{2}c^{3}}{k}+\xi _{\gamma }\frac{%
k^{2}}{2E_{0}c}
\end{equation*}
When $\xi _{\gamma }<0$ \ \ $q_{\shortparallel }^{\min }$ becomes negative
at the critical energy $k_{cr}=[\frac{4(mc^{2})^{2}E_{0}}{\xi _{\gamma }}]^{%
\frac{1}{3}}\simeq 2.2$x$10^{13}$ eV for $\mid \xi _{\gamma }\mid =1$.

Let us now calculate the \emph{suppression factor S}, that measures the
relative change of the interaction cross section and is defined as: 
\begin{equation}
S=\frac{\frac{d\sigma }{dE}}{\frac{d\sigma }{dE_{BH}}}=\frac{l_{f}}{l_{0}}
\end{equation}
where $d\sigma /dE$ is the differential cross section with suppression, $%
d\sigma /dE_{BH}$ is the Bethe-Heitler (BH) cross section, and $l_{f}$ is
the formation length with suppression. This definition is convenient for
estimation of suppression due to different factors because it is easy to
estimate the change of the quantity $q_{\shortparallel }$ caused by these
factors and, respectively $l_{f}$. For example, for strong suppression by
multiple scattering, $S_{LPM}=\sqrt{\frac{E_{LPM}k}{E(k-E)}}$ (the material
dependent energy $E_{LPM}$ is defined in \cite{Stanev}). For $E\approx
k-E\approx k/2$ \ \ \ $S_{LPM}\simeq 2\sqrt{\frac{E_{LPM}}{k}\text{.}}$

In case of pair production LIV gives a suppression factor 
\begin{equation}
S_{LIV}=\frac{1}{1+\xi _{\gamma }\frac{k^{3}u(1-u)}{E_{0}(mc^{2})^{2}}}
\end{equation}
(here $u=E/k$) and for $E\approx k-E\approx k/2$ \ 

\begin{equation}
S_{LIV}\simeq \frac{1}{1+\xi _{\gamma }\frac{k^{3}}{4E_{0}(mc^{2})^{2}}}
\end{equation}
Neglecting $1$ in the denominator for strong suppression, Eq.10 becomes 
\begin{equation}
S_{LIV}\simeq \frac{4E_{0}(mc^{2})^{2}}{\xi _{\gamma }k^{3}}
\end{equation}

Some numerical values for $S_{LIV}$ \ for pair production ($E_{0}=10^{28}$
eV , $\xi _{\gamma }=1$ ) are shown in table~\ref{tab1}. For comparison LPM
suppression factors for air ($E_{LPM}$=2.34$\times $10$^{8}$ GeV at sea
level) and lead ($E_{LPM}$=4.3$\times $10$^{3}$ GeV) are also shown in the
table (the values for $E_{LPM}$ are taken from \cite{Klein}). 
\begin{table}[tbp]
\caption{Suppression factors for pair production}
\label{tab1}
\begin{center}
\begin{tabular}{|l|lll|}
\hline
$k(TeV)$ & $S_{LIV}$ & $S_{LPM}^{air}$ & $S_{LPM}^{Pb}$ \\ \hline
$1$ & $1$ & $1$ & $1$ \\ 
$10$ & $0.91$ & $1$ & $1$ \\ 
$100$ & $10^{-2}$ & $1$ & $0.41$ \\ 
$1000$ & $10^{-5}$ & $1$ & $0.13$ \\ \hline
\end{tabular}
\end{center}
\end{table}

If $\xi _{\gamma }=-1$ \ \ \ \ 
\begin{equation*}
S_{LIV}=\frac{1}{1-\frac{k^{3}}{4E_{0}(mc^{2})^{2}}}
\end{equation*}
$S_{LIV}$ turns into enhancement, which becomes infinite for $%
k=k_{thr}\equiv\lbrack 4E_{0}(mc^{2})^{2}]^{\frac{1}{3}}\approx 2.2 \times
10^{13} $ eV$\equiv 22$ TeV, as shown in table~\ref{tab2}.

\begin{table}[tbp]
\caption{Enhancement factor for pair production}
\label{tab2}
\begin{center}
\begin{tabular}{|l|l|}
\hline
$k(TeV)$ & $S_{LIV}$ \\ \hline
$1$ & $1$ \\ 
$10$ & $1.11$ \\ 
$15$ & $1.2$ \\ 
$20$ & $4.27$ \\ \hline
\end{tabular}
\end{center}
\end{table}

This means that when $\xi $ is negative the formation length of the process
(pair production in this case), respectively the cross section, increases.
This sharp increase around $k_{thr}$ could in some cases be compensated by
multiple scattering and other suppression factors.

The effects of LIV on bremsstrahlung are similar to those for pair
production. If we neglect the last term in (6) (we suppose that LIV is
negligible for low energy bremsstrahlung photons) the suppression factor
becomes 
\begin{equation}
S_{LIV}=\frac{1}{1+\xi _{e}\frac{2E^{3}(1-u)}{E_{0}(mc^{2})^{2}}}
\end{equation}
or, for strong suppression, 
\begin{equation}
S_{LIV}\simeq \frac{E_{0}(mc^{2})^{2}}{2\xi _{e}E^{3}(1-u)}
\end{equation}

The corresponding LPM suppression factor (also for strong suppression) is 
\begin{equation}
S_{LPM}=\sqrt{\frac{E_{LPM}}{E}\frac{u}{1-u}}
\end{equation}
Some numerical values for $\xi =1$ and $u=0.001$ are compared in table~\ref
{tab3}. 
\begin{table}[tbp]
\caption{Suppression factors for bremsstrahlung}
\label{tab3}
\begin{center}
\begin{tabular}{|l|lll|}
\hline
$E(TeV)$ & $S_{LIV}$ & $S_{LPM}^{air}$ & $S_{LPM}^{Pb}$ \\ \hline
$1$ & $1$ & $1$ & $6.6$x$10^{-2}$ \\ 
$10$ & $0.57$ & $1$ & $2.1$x$10^{-2}$ \\ 
$100$ & $1.3$x$10^{-3}$ & $1$ & $6.6$x$10^{-3}$ \\ 
$1000$ & $1.3$x$10^{-6}$ & $0.48$ & $2.1$x$10^{-3}$ \\ \hline
\end{tabular}
\end{center}
\end{table}

\section{Discussion and conclusions}

Using the concept of formation length we have shown that LIV parameters that
are necessary for the explanation of the existence of UHECR and the
non-absorption of 20 TeV $\gamma$-rays in a particular Lorentz invariance
violation model will also strongly affect the bremsstrahlung and pair
production cross sections. The general form of Planck scale motivated LIV
dispersion relations is of the form 
\begin{equation*}
E^{2}-p^{2}-m^{2}\simeq \eta E^{2}( \frac{E }{E_{0}})^{\alpha }\simeq \eta
p^{2}(\frac{E}{E_{0}})^{\alpha} \; ,
\end{equation*}
where $\alpha $ and $\eta $ are free parameters. We have only analyzed the
case $\alpha =1$ and $\mid \eta \mid =1$.

To explain the experimental astrophysical paradoxes the positive values of $%
\eta$ should be excluded~\cite{Amelino}, which means that the parameter $\xi 
$ should be positive and the LIV would suppress pair production and
bremsstrahlung above some critical energy $k_{cr}$. We obtain $k_{cr}$
values that are of the same order of magnitude as the critical energy $E_{c} 
$ defined in \cite{Kifune} (see also \cite{Sigl}). In the case of LIV $E_c$
is obtained from the condition for a minimum target photon energy for pair
creation in a soft photon field. Above $E_{c}$ the target photon energy
grows (for $\xi >0 $) until the Universe becomes transparent to ultrahigh
energy photons . This reflects the fact that (\cite{Aloisio,Liberati}) one
can expect deviations from standard kinematics when the last two terms in
the dispersion relation $E^{2}\approx p^{2}+m^{2}+\xi p^{3}/E_{0}$ are of
comparable magnitude. For $\xi =1$ the condition becomes $p_{dev}\sim
(m^{2}E_{0})^{\frac{1}{3}}\sim 10$ TeV.

The suppression factor $S$ calculated above and shown for pair production in
Table~\ref{tab1} is very strong and does not depend on the target material
as in LPM. The suppression increases very fast with energy,
proportionally to $E^{3}$ above 100 TeV. This will make photons and
electrons very penetrating particles and will drastically suppress the
electromagnetic shower development. For example, only about 20\% of 100 TeV
primary photons will interact in the atmosphere. For $\gtrsim 300\div 400$
TeV photons the atmosphere will be transparent. Such an effect must have
already been observed in the numerous experiments in cosmic rays. The depth
of maximum in electromagnetic showers $X_{max}$ is proportional to the
product of the radiation length $X_{0}$ and the logarithm of the primary
energy $\ln {E}$. In the case of LIV the radiation length $X_{0}$ above the
critical energy becomes infinite, which changes drastically the behavior of
electromagnetic and of hadronic air showers. This is especially true for
giant air showers where the depth of maximum $X_{max}$ is generally
determined by the electromagnetic cascades of primary energy exceeding 10$%
^{17}$ eV. Thus LIV with the parameters discussed above would contradict to
the results from giant air showers~\cite{NagWat}.

Coleman and Glashow \cite{Coleman} have suggested a different scheme for LIV
in which the maximum attainable velocity $c_{a}$ of a particle is different
from the photon velocity $c$. The relevant dispersion relations then has the
form $E_{a}^{2}=p^{2}c_{a}^{2}+m_{a}^{2}c_{a}^{4}$. In this case our results
will be applicable by substituting $c^{2}-c_{a}^{2}$ for $\xi \frac{E}{E_{0}}
$ in (1). Then the critical energy will be defined as $k_{cr}=\frac{\sqrt{8}%
m_{e}}{\mid c^{2}-c_{e}^{2}\mid ^{1/2}}$ . If the shower development is
observed with no deviation from the standard cascade theory up to $\simeq 22$
TeV, this will put the limits $E_{0}\gtrsim 10^{28}$ eV (Planck mass) or $%
\mid c^{2}-c_{e}^{2}\mid \lesssim 1.5 \times 10^{-15}$.

The observations of giant atmospheric showers created by particles with
energies $\gtrsim $ $10^{20}$ eV give, in principle, the possibility for a
more precise test of LIV. For example, if the case $\alpha =1$ must be ruled
out, one can move to the $\alpha =2$ case, i.e. quadratic suppression of $%
E_{0}$. The suppression factor for pair creation then becomes $S_{LIV}\simeq 
\frac{4E_{0}^{2}(mc^{2})^{2}}{\xi k^{4}}$ and the drastic deformation of the
shower development would be observed at energies $\gtrsim 10^{17}$ eV. In
the frame of Coleman and Glashow scheme this will put the limits $\mid
c^{2}-c_{e}^{2}\mid \lesssim 10^{-23}$. In this connection we would like to
point out the interesting work \cite{Dedenko} where new constraints on $\mid
c_{\gamma }-c_{\pi ^{0}}\mid $ ($c_{\pi ^{0}}$ is the maximal attainable
speed of neutral pions) are obtained by comparing the experimentally
measured position of the shower maximum $X_{\max }$ with calculations.

Our estimates are based on a quite general quantum-gravity induced
modification of the dispersion relation between the energy and the momentum
of a particle. We also supposed that the sign of the free parameter $\xi $
is fixed. As noted above, the positive sign of $\xi $ is required to solve
the astrophysical paradoxes and is consistent with the constraints deduced
from the analysis in \cite{Liberati}. The future studies using the fruitful
and physically very clear concept of the formation (coherent) length could
include some additional phenomenological suggestions made in the literature.
For example, it is possible to construct schemes in which the classical
relation $E^{2}=p^{2}+m^{2}$ holds only on average, but in a given physical
realization $E^{2}=p^{2}+m^{2}+\Delta $, with $-\xi p^{2}(\frac{E}{E_{0}}%
)^{\alpha }<\Delta <\xi p^{2}(\frac{E}{E_{0}})^{\alpha }$ \cite{Ng}. It is
also possible that energy conservation (assumed in this work) may be valid
only in a statistical sense (see the review \cite{Subir} and references
therein). If the LIV reflects some kind of special property of the
space-time at Planck scales (or some other scale) the sign of $\xi $ may
fluctuate on the length scales of the order of the Planck length. The
negative sign of $\xi $ will, however, complicate the analysis. We showed
above that for negative $\xi $ the formation length increases with energy,
becoming infinite at the critical energy and negative above it. It is not
obvious how to interpret the infinite, or negative, formation length in
interactions with the field of a nucleus.


Finally, if the LIV is obtained as a result of spontaneous symmetry breaking
in the context of an explicitly Lorentz invariant theory (see, e.g. \cite
{Kost}), it is possible that another scale (light mass scale $m_{l}$) enters
the calculations. The correction term for the modified dispersion relation
in this case is of the form (see (1)) $\xi \frac{E^{2}V}{M_{P}c^{2}}$, where
V is vev after the symmetry breaking and $M_{P}$ might be the Planck mass.
In this case we could expect much smaller LIV depending on the value of $%
m_{l}$. For example, if $V$ is of the order of the other fundamental energy
scale in nature, the electroweak scale $m_{EW} \sim 10^{3}$ GeV, then one
can expect the deviation from the standard kinematics above $p_{dev}\sim 50$
TeV. The corresponding $S_{LIV}$ for pair production will be 0.51 for 100
TeV and 0.01 for 1000 TeV (instead of corresponding values $10^{-2}$ and $%
10^{-5}$ given in Table 1).

Acknowledgments. The work of TS is supported in part by the US Department of
energy contract DE-FG02 91ER 40626.

\end{document}